# ULV: A robust statistical method for clustered data, with applications to multi-subject, single-cell omics data

Mingyu Du[1*], Kevin Johnston[2*], Veronica Berrocal[3], Wei Li[4], Xiangmin Xu[2,5#], Zhaoxia Yu[3,5#]

[1]Center for Complex Biological Systems, University of California, Irvine, 92697, CA, USA.
[2]Department of Anatomy and Neurobiology, University of California, Irvine, 92697, CA, USA
[3]Department of Statistics, University of California, Irvine, 92697, CA, USA.
[4]Division of Computational Biomedicine, Department of Biological Chemistry, School of Medicine, University of California, Irvine, 92697, CA, USA.
[5]Center for Neural Circuits Mapping, University of California, Irvine, 92697, CA, USA.

*Co-first authors: mingyd1@uci.edu; kgjohnst@uci.edu
[#]Corresponding author(s). E-mail(s): xiangmix@hs.uci.edu; zhaoxia@ics.uci.edu;
Contributing authors: vberroca@uci.edu; wei.li@uci.edu;

## Abstract

Molecular and genomic technological advancements have greatly enhanced our understanding of biological processes by allowing us to quantify key biological variables such as gene expression, protein levels, and microbiome compositions. These breakthroughs have enabled us to achieve increasingly higher levels of resolution in our measurements, exemplified by our ability to comprehensively profile biological information at the single-cell level. However, the analysis of such data faces several critical challenges: limited number of individuals, non-normality, potential dropouts, outliers, and repeated measurements from the same individual. In this article, we propose a novel method, which we call U-statistic based latent variable (ULV). Our proposed method takes advantage of the robustness of rank-based statistics and exploits the statistical efficiency of parametric methods for small sample sizes. It is a computationally feasible framework that addresses all the issues mentioned above simultaneously. We show that our method controls false positives at desired significance levels. An additional advantage of ULV is its flexibility in modeling various types of single-cell data, including both RNA and protein abundance. The usefulness of our method is demonstrated in two studies: a single-cell proteomics study of acute myelogenous leukemia (AML) and a single-cell RNA study of COVID-19 symptoms. In the AML study, ULV successfully identified differentially expressed proteins that would have been missed by the pseudobulk version of the Wilcoxon rank-sum test. In the COVID-19 study, ULV identified genes associated with covariates such as age and gender, and genes that would be missed without adjusting for covariates. The differentially expressed genes identified by our method are less biased toward genes with high expression levels. Furthermore, ULV identified additional gene pathways likely contributing to the mechanisms of COVID-19 severity.

## Introduction

Single-cell omics technology has revolutionized our ability to conduct biological research, enabling measurements of gene expression and protein levels across an unprecedented number of cells. This technology allows for the quantification of these modalities not just at the aggregate level, but at the single-cell resolution, offering a detailed view of cellular heterogeneity. Consequently, researchers can construct comprehensive cell-type atlases of various tissues and organs, providing deeper insights into the complex symphony of cellular activities and interactions.

Accompanying the technological advancements, methodological developments have also been fast evolving and have played critical roles in drawing biologically relevant conclusions. Single-cell RNA sequencing profiles, which measure the transcriptomes of individual cells, are inherently confounded and clustered by multiple factors such as subject and batch. It is increasingly understood that such data clustering structures introduce dependencies, as observations from the same cluster tend to be more correlated than observations from different ones. Numerous studies, including our own work, have documented that false conclusions can be drawn when ignoring data dependencies due to clustering [1–3]. Several recent papers reported that naively applying popular methods developed for bulk sequencing data on single-cell RNA sequencing data (Fig. 1) can result in large numbers of false positives [4–7]. To mitigate such bias, two strategies have been suggested. The first approach, often known as "pseudobulk", constructs independent observations from single-cell sequencing data by aggregating cells within each biological unit. The other strategy is to use a mixed-effects model in which the parameters of interest are treated as fixed effects while the dependence in the single-cell data is explicitly modeled by introducing subject-level random effects in regression frameworks. Pseudobulk is computationally very efficient as normalized and aggregated gene counts can be analyzed using methods for bulk sequencing [8–10]. On the other hand, mixed-effects models are often computationally demanding and might not sufficiently account for batch effects in large scale experiments.

One additional and significant challenge in single-cell measurements is that single-cell profiles are typically not normally distributed. For single-cell RNA-seq data, this problem has been partially addressed via parametric methods that use discrete distributions such as negative binomial, Poisson, or a mixture of a point mass at zero and a parametric distribution [8–10]. Note that these methods cannot be applied to single-cell proteomics data as protein levels are continuous. Alternatively, nonparametric methods, such as the rank sum test, also known as the Wilcoxon-Mann-Whitney test, have been advocated as robust alternatives to parametric methods due to their relaxed assumptions on distributions. One study compared methods for analyzing bulk sequencing data and concluded that the Wilcoxon rank sum test is the best-performing one for studies with sufficient sample sizes [11].

For single-cell data, the Wilcoxon rank sum test is the most widely used test; however, it is mostly implemented to pooled cell data by treating cells as independent units. This leads to false positives [5]. Conversely, while using the rank sum test on subject-level summary data is recommended [11], its limitations become apparent when considering the minimum attainable p-value in relation to the sample size. Specifically, the minimum attainable p-value of subject-level Wilcoxon rank sum test based on *m* cases and *n* controls is $1/\binom{m+n}{m}$. For example, with $m = n = 5$, the minimum p-value is approximately 0.004, making it impossible to detect differentially expressed genes in large-scale studies due to the stringent p-value cutoff required to adjust for multiple comparisons across large numbers of genes.

In this article, we introduce the U-statistic based Latent Variable (ULV) model for the analysis of single-cell data. This method is designed to simultaneously address the three major challenges present in single-cell data: small sample size, non-normality, and clustering. Our innovative approach combines rank-based statistics with a latent variable model, providing a robust solution capable of handling distributional violations and dependencies caused by clustering. It is versatile in modeling different types of single-cell data. Moreover, it offers flexibility in adjusting for covariates, thereby mitigating potential bias due to technically and scientifically relevant confounders. We benchmark our method against six competing methods using simulated datasets. We also apply it to two studies: one involving single-cell proteomics of acute myelogenous leukemia (AML) and another focusing on single-cell RNA sequencing of COVID-19 symptoms.

## Results

### *Overview of the U-statistic based Latent Variable (ULV) model*

Our U-statistic based latent variable model (ULV) builds upon a generalized version of the Wilcoxon rank sum test, which is also known as the Mann–Whitney U test (Fig. 2). Briefly, ULV is a two-stage robust statistical framework to evaluate whether cases and controls show differential expression. In the first stage, ULV utilizes a nonparametric rank-based method to compute a difference measure between each pair of individuals from distinct groups (Fig. 2b). In the second stage, ULV postulates a latent level for each subject via a parametric model (Fig. 2c).

Stage 1 of the framework relies on a choice of difference metric between case subjects and control subjects. This framework covers several pseduobulk methods as special cases. For example, when the difference is calculated using the mean/median expression of two subjects, the test corresponds to performing a two-sample t-test using mean/median as the subject level summary; when the difference is whether the mean/median expression level of one subject is greater than another one, the test is equivalent to the one-sided Wilcoxon rank sum test with the mean/median as the summary quantity for each subject. We recommend using the probability index (PI) as the difference measurement between two subjects because it is robust to outliers. In layman's terms, the PI between a disease subject and a healthy one quantifies how the cells from one subject are separated from those of the other. Statistically, the PI estimates the probability that a randomly selected cell from the diseased subject has higher expression level than that of a randomly selected cell from a healthy subject. Mathematically, it is the rescaled version of the Wilcoxon rank sum statistic or the equivalent U statistic. Finally, from a classification point of view, the PI is related to the receiver operating characteristic curve (ROC, see methods): the PI is the area under the curve (AUC) when using a gene to discriminate between the cells of two subjects (Fig. 2b).

Having calculated the pairwise differences in Stage 1, our ULV method models them using a latent variable model. By adopting the PI as a difference metric, our ULV method enjoys the advantages of both parametric and nonparametric methods. On one side, the rank-based method of the first stage ensures that the analysis is robust to outliers and non-normal distributions of the data, on the other, the latent variable model of the second stage efficiently accounts for correlations in the pairwise between-subject differences. In the absence of any covariate, $d_{ij}$, the difference between the $i$th disease subject and the $j$th control subject, is modeled as the difference of their latent expression levels, denoted by $a_i$ and $b_j$, respectively. The latent variables $a_i$ and $b_j$ are in turn assumed to follow normal distributions and they are introduced for two reasons. First, the difference in their means, denoted by $\mu$, is the parameter of interest, as it represents the mean difference between the case and the control groups at the population level. Second, they also explicitly model the dependence between the pairwise differences. For example, the dependence between $d_{12}$ and $d_{13}$ is ensured because they share the latent variable $a_1$. Additionally, our ULV model is very flexible in that: (i) it offers covariate adjustment; (ii) it can be easily extended from two-group comparisons to multi-group comparisons; and (iii) it provides the option of a weighted analysis to accommodate clusters of varying sizes (Fig. 2c). ULV is available as an R package at https://github.com/yu-zhaoxia/ULV.

*ULV has better control of type I error rates with comparable power to existing methods*

We conducted simulation studies to assess the performance of the ULV method, and we benchmarked it against six existing methods, including three popular differential expression analysis methods (DESeq2, MAST, and Wilcoxon rank sum test). For DESeq2 and Wilcoxon rank sum, we assessed their performance both using single-cell (pooled cell) data and at the subject (pseudobulk) level, which leads to four tests: DESeq2_sc, DESeq2_pseudobulk, Wilcoxon_sc, and Wilcoxon_pseudobulk. For the MAST method, we examined the performance using generalized linear models (GLM) or mixed effects models (i.e. MAST_glm and MAST_mixed). A more detailed presentation and discussion of the methods compared as well as more information on the simulation setting are provided in the Methods section. Briefly, we adopted the simulation strategy utilized in the BSDE method [7], which simulates gene expression count data at single-cell level using a zero-inflated negative binomial distribution with parameters estimated from real-world scRNA-seq data. For each subject, we generated gene expression data for 1000 genes. We considered various settings with respect to the total number of subjects per condition, the total number of cells per subject, covariate adjustment, and weighted analysis to account for varying numbers of cells. For each setting, 100 simulations were conducted, and the boxplots of the type I error rates or power are presented in Fig. 3.

As expected, the cell-level tests, namely DESeq2_sc, MAST_glm, and Wilcoxon_sc, all treated cells as independent units, resulting in extremely inflated type I error rates for each of the four significance levels (0.001, 0.01, 0.05, and 0.2) (Fig. 3a). Note that their type I error rates exceed the limit of the y-axis when the significance level is 0.01 or 0.001; thus, they are only partially shown in these panels.

Consistent with findings reported by others [11], although pseudobulk methods significantly reduce type I error rates compared to their cell-level counterparts, they still exhibit moderate inflation in the number of false positives. Fig. 3 indicates that both DESeq2_pseudobulk and MAST_mixed showed varying degrees of inflation in type I error rates.

Among all the methods tested, ULV has the best control of false positives since it maintains valid type I error rates across all simulation settings. ULV has the desired type I error rate at all significance levels and for all cluster sizes (Fig. 3a). Conversely, although Wilcoxon_pseudobulk controls type I error rates, but is consistently conservative, as illustrated by the small type I error rates compared to the corresponding chosen significance levels. This conservatism arises because Wilcoxon test at the subject level generates a minimum p-value of $1/\binom{m+n}{m}$ when the data consist of *m* cases and *n* controls. Thus, when using more stringent significance levels or when sample sizes are smaller, it becomes challenging for Wilcoxon_pseudobulk test to identify DE genes.

We next examined two extensions of the ULV method: ULV_adj (adjusted), which adjusts for covariates and ULV_wt (weighted), which accounts for varying cluster sizes. In scenarios involving a covariate, ULV_adj is the only method achieving the desired type I error rate (Fig. 3b); importantly, it remains valid even in the absence of covariate effects. When the cluster sizes are unbalanced, both ULV and its weighted version ULV_wt have the best control of type I error rates, although the weighted version shows a slight inflation of type I error rates (Fig. 3c).

The simulation results also suggest that ULV has satisfactory statistical power (Fig. 3 d,e). The power of all tests increases with sample size (from $n$ = 5 to $n$ = 10) and fold change (from $r$ = 1.5 to $r$ = 2). ULV and ULV_wt exhibit slightly lower but comparable power to DESeq2_pseudobulk and MAST_mixed. The tests based on cell-level analysis have the greatest power; however, we do not recommend using these methods for differential expression analysis due to their extremely inflated type I error rates (Fig. 3a-c). As expected, the subject-level Wilcoxon test consistently exhibits the lowest power across all scenarios.

### *ULV discovers differentially expressed proteins in acute myelogenous leukemia patients*

Here we demonstrate the usefulness of our ULV method when applied to multi-subject single-cell proteomics data from a recent study of acute myelogenous leukemia (AML) [12]. This study examined bone marrow protein expression levels in AML patients and healthy controls, focusing on malignant and healthy hematopoietic stem and progenitor cells (HSPCs) as well as myeloid cells from pediatric subjects. For data quality control, we included only those subjects with more than 50 cells and proteins expressed in more than 10% of the cells from AML patients. This resulted in a dataset comprising 269 proteins from 56,764 HSPC cells from 16 pediatric AML patients and 5 age-matched controls, and 270 proteins from 65,510 myeloid cells from 13 pediatric AML patients and 5 age-matched controls. The raw protein expression data was pre-processed using centered log-ratio normalization. Our objective is to identify proteins that are differentially expressed in HSPCs or myeloid cells between pediatric AML patients and age-matched controls.

Since the cell number varies substantially from patient to patient (Fig. 4a), we used ULV_wt, the weighted version, to account for unbalanced cell numbers. A protein was considered differentially expressed if (1) the FDR-adjusted p-value is below 0.1 and (2) its probabilistic index is lower than 0.45 or greater than 0.55. Using these criteria, ULV_wt detected 6 differentially expressed proteins between AML pediatric patients and controls in HSPC cells and 19 in myeloid cells (Fig. 4b).

A surface immune receptor, CD244-2B4, was identified as a protein over expressed in both HSPC and myeloid malignant cells in pediatric AML patients. Increased CD244 expression in tissue cells from AML patients has been previously reported to increase the proliferation of leukemia cell lines [13], which further suggests that CD244 is an AML-associated biomarker. The upregulation of another identified protein in both cell types, programmed cell death protein 1 (CD279-PD-1), has been linked to the suppression of immune responses during tumor development [14] and is also associated with AML pathology [15] (Fig. 4c). Similarly, a significant increase of a CXC chemokine receptor, CD183-CXCR3, was also detected in both cell types in AML pediatric patients. CD183-CXCR3 plays an important role in regulating the function of the immune system. Its expression levels are closely associated with AML's oncogenic processes [16]. Two proteins showing overexpression in AML patients in HSPC cells only are the interleukin 3 receptor alpha (CD123, Fig.4c) and a transmembrane phosphoglycoprotein protein (CD34). These two proteins have been previously identified: CD123 was found as a putative marker in AML blasts compared with normal HSPCs [17] and CD34 is a well-known biomarker for identifying malignant stem cells in leukemia and other cancers [18].

As a comparison, Wilcoxon_pseudobulk test failed to detect any differentially expressed proteins. Three options of subject-level summary statistics were considered: mean, median, and the proportion of non-zero expression values. None of these tests identified significantly differential proteins at the false discovery rate (FDR) of 0.1. This is consistent with what we found in the simulation study: the pseudobulk version of the Wilcoxon test has low statistical power when sample sizes are small.

Lastly, to verify that ULV maintains valid type I error rates, we generated 100 null datasets by permuting the disease condition labels of pediatric AML patients and their age-matched healthy controls. These permutations ensure that all identified DE proteins are false positives. We evaluated the type I error rate for three different DE methods (ULV_wt, Wilcoxon_sc, and Wilcoxon_pseudobulk) by applying them to the permuted datasets. In alignment with the findings from the simulation studies, Wilcoxon_sc exhibits extremely inflated type I error rate across various p-value thresholds, whereas Wilcoxon_pseudobulk is overly conservative (Fig. 4d). In contrast, only ULV_wt shows type I error rates that align closely with the nominal significance levels across all the different settings considered.

*ULV accurately detects differentially expressed genes dependent on disease severity in a COVID-19 scRNA-seq data after adjusting for covariates*

Next we analyzed scRNA-seq profiles in peripheral blood mononuclear cells (PBMC) from patients with mild or severe symptoms of Coronavirus disease 2019 (COVID-19) [19]. Gene expression at single-cell resolution was obtained using a droplet-based single-cell platform (10x Chromium). The Unique Molecular Identifier (UMI) count matrices, comprising 46,584 genes and 99,049 cells, were imported into our analysis pipeline. For data quality control, we filtered out genes expressed in fewer than 10% of cells to remove low-abundance transcripts. Additionally, pooling normalization was applied to remove cell-specific biases introduced by varying read depths and other technical issues during the sequencing process [20]. Several recent studies have indicated that severe COVID-19 cases often involve the exhaustion of CD8+ T cells ([21], [22]), making CD8+ T cell a particularly intriguing cell type for further investigation. For this reason, we focused on CD8+ T cells. Patients with fewer than 50 CD8+ T cells were excluded. Consequently, our refined dataset consisted of 5,159 genes from 5,494 cells across 15 patients, including 5 with mild COVID-19 and 10 with severe COVID-19.

In addition to severity of COVID-19, information on gender and age of each patient is also available. To account for potential confounders due to these two covariates and the varying cell counts of patients, we analyzed the data using the ULV_wt method and we adjusted for either age or gender (Fig. 5b-c). To understand how covariates can affect gene expression, we visualize two example genes. EIF1AX (Eukaryotic translation initiation factor 1A X-linked) encodes the host protein involved in the COVID-19 viral RNA/protein synthesis processes by enhancing ribosome dissociation and stabilizing the binding of the initiator Met-tRNA to the 40S ribosomal subunit [23, 24]. It displays gender-dependent expression levels (Fig. 5b). This gene was not identified as differential expressed (FDR adjusted p-value = 0.85) when the model did not adjust for gender but it became significant (FDR-adjusted p-value = 0.05) when we adjusted for gender. Consistent with what we found, a recent study [25] reported that EIF1AX displays marked difference between females and males in regulatory T cells.

Another example gene is ENO1 ($\alpha$-enolase), which encodes a glycolytic protein. The gene shows age-dependent expression (Fig. 5c) and was not identified as differentially expressed (FDR-adjusted p-value = 0.20) by the ULV method without age adjustment but it was identified as upregulated (FDR adjusted p-value = 0.03) after adjusting for age. The pairwise difference in gene expression increases with the age difference between severe and mild patients (Fig. 5c). A recent study also revealed that ENO1 has increased expression level in peripheral blood neutrophil proteomes in COVID patients [26]. These results indicate that the ULV method with covariate adjustment can identify scientifically and biologically relevant signals, and thus may offer more information toward disease mechanisms. Overall, we found that the ULV method with covariate adjustment identified 137 genes as differentially expressed that were missed by the basic ULV method (Fig. 5d).

In DE analysis using scRNA-seq data, one concern is potential bias toward genes with high expression levels [5]. To examine whether we encounter similar issues in the COVID-19 data, we visualize the mean expression level and the proportion of genes with zero expression measurements for the 100 top-ranked genes identified by each DE methods (Fig. 6a-b). Both age and gender effects were adjusted in all DE methods except for the two Wilcoxon tests. Genes detected by methods that treat cells as individual samples have higher mean expression levels and lower proportions of genes with zero expression measurements than the other DE methods. The ULV_wt method (including covariate adjustment) is one of the two methods that were least biased by gene expression levels.

An important question in DE analysis is identifying biologically relevant pathways. We conducted gene set enrichment analysis (GSEA) using the signed ranks of genes based on DE analysis from DESeq2_pseudobulk and ULV_wt, which identified 207 and 158 pathways respectively, with 129 pathways detected by both methods (Fig. 6c). Among the enriched pathways that were detected by ULV_wt but not by DESeq2_pseudobulk (Fig. 6d), the most significant is the EPH (Erythropoietin Producing Human Hepatocellular)-Ephrin related pathway. EPH-Ephrin signaling regulates cell-cell adhesion and cell-matrix adhesion, crucial for maintaining tissue architecture and function. Alterations in cell adhesion and migration can affect how the COVID-19 virus spreads within tissues. Recent studies [27, 28] have shown that EPH receptors could serve as potential entry points for the COVID-19 virus, facilitating contact and stimulating downstream signaling pathways. This interaction may contribute to COVID-19 disease progression by affecting cellular functions such as adhesion, proliferation, differentiation, and migration.

Several interleukin (IL)-related pathways were also identified by ULV_wt but not by DESeq2_pseudobulk (Fig. 6c-d). ILs are a group of cytokines that play crucial roles in regulating immune responses. Previous studies suggest that ILs are related to the development of COVID-19 [29] and increased IL levels are associated with COVID-19 disease severity [30]. Additionally, a pathway related to Fc-Gamma receptor dependent phagocytosis was also enriched by genes identified by ULV_wt. Fc-Gamma receptor dependent phagocytosis plays a crucial role in the immune response to COVID-19 [31] by facilitating clearance of the COVID-19 virus through antibody-dependent cellular phagocytosis, where antibodies bind to the virus, allowing phagocytes to ingest and destroy it. Summarizing all the results, ULV_wt allowed the identification of several notable pathways that DESeq2_pseudobulk did not detect, demonstrating the potential of our new method to uncover significant biological insights.

## Discussion

Motivated by the challenges in analyzing single-cell data, we have developed a new robust method for the analysis of clustered data. By combining rank statistics with a latent variable model, our method is not only robust to outliers and violations of distributional assumptions, but also achieves substantially greater power compared to conventional rank tests based on cluster-level summary data. Our method provides an important resource that straddles the middle ground between pseudobulk and single-cell differential expression approaches by utilizing a statistic that quantifies differences between sample pairs. Our method is also less restrictive in terms of assumptions than are commonly needed by parametric methods, and thus it is usable for other types of single-cell data such as proteomics.

Our simulation studies have confirmed that the ULV method maintains consistent type I error rates across a wide range of scenarios. We found that, at all significance levels, the ULV method exhibited lower type 1 error rates than all others single-cell based methods considered, and lower rates than those of all pseudobulk methods except for Wilcoxon_pseudobulk. Additionally, the power of the ULV method was 2-3 times higher than that of Wilcoxon_pseudobulk and comparable to that of all the other pseudobulk methods.

We next applied our methodology to single-cell proteomic and transcriptomic datasets. For the proteomic study, ULV identified critical protein regulations that were not singled out using approaches based on Wilcoxon test. Similarly, in the scRNA-seq study we demonstrated that ULV successfully accounts for covariate-induced biases (age and gender). Compared with alternative methods, we showed that the ULV method exhibits very little (if any) bias toward highly expressing genes, and it is able to pinpoint critical regulatory pathways missed by DESeq2_pseudobulk.

One advantage of our method is its ability to effectively mitigate bias from potential confounding factors by accounting for the effects of covariates. This is achieved because the latent variable model stage of the ULV method operates within a regression framework, which allows to naturally adjust for covariates. Inferring causal relationships, rather than merely testing for associations, has gained growing interest in recent decades. For causal inference, additional tools beyond simply adding covariates are necessary. In our future research, we plan to integrate weighted or propensity score methods into our framework to further reduce bias in observational studies.

The rank-based statistic we chose is the estimated probabilistic index between two clusters. This is a reasonable choice due to its connection with AUC and the rank sum statistic. One cautionary note is that the concept of "probabilistic greater" is not transitive, as exemplified by intransitive dice. In situations involving multiple treatment groups, where the focus is on comparing the difference between two specific groups, we recommend directly comparing them, rather than assessing their difference by comparing both to a common reference group. An alternative solution is to consider other statistics, such as the Hodges-Lehmann statistic [32, 33], which offers a robust estimate of the location of the difference between two populations. Furthermore, testing for unequal variance is relevant in certain situations. While our framework can be adapted to construct such tests by comparing the variances of latent variables, we acknowledge that this area requires further investigation and potential application exploration. These topics will be part of the focus of our future work.

Overall, we believe that this work provides a critical resource for biologists and bioinformaticians analyzing biological features at the single-cell level. We believe that the ULV method contributes to biological progress because of its control of type I error rates combined with its flexibility for use across multiple data modalities.

## Acknowledgements

The authors would like to thank Dr. Eran Mukamel for invaluable discussions and suggestions that greatly contributed to the development and completion of this project. This work was supported by National Institutes of Health (NIH) grants U01AG076791, U01DA052769, R01AG067153, R01AG082127 and RF1AG065675 to X.X. K.G.J. acknowledges support from NIH grant T32 DC010775-14.

**Author contributions:** Z.Y., M.D. and K.G.J conceptualized the work. Z.Y., M.D. and K.G.J. wrote the manuscript. X.X. oversaw and supported the work. W.L. and V.B. advised and collaborated on the methods presented here and edited the manuscript.

## Ethics Declaration

All authors have no competing interests.

## Methods

### *Revisiting the Wilcoxon-Mann-Whitney test*

Before presenting the ULV method, it is necessary to provide the background of the Wilcoxon rank sum test and related U statistics. This is because our ULV is an extended version of the Wilcoxon rank sum test and the difference metric we use to compare a case and a control is itself related to the Wilcoxon rank sum test statistic based on the cells of two subjects. In two-sample comparison problems, the conventional t-test compares the means of the two underlying populations from which the samples are drawn. However, when dealing with small sample sizes, the Wilcoxon rank sum test [34] is often preferred. This is because the two-sample t-test relies on normality assumptions or large-sample approximations, which may not be appropriate in small samples. The Wilcoxon rank sum test, as suggested by its name, replaces the original outcome values with their ranks, offering robustness against outliers and other distributional assumption violations. Notably, this test is also referred to as the Wilcoxon-Mann-Whitney test due to its equivalence to the Mann-Whitney U Test [35], which is shown below.

Consider two independent samples, where $(y_{11},\cdots,y_{1m})$ is a random sample from the distribution of the cases, $F_1$ and $(y_{01},\cdots,y_{0n})$ is a random sample from distribution $F_0$. Define $R_{1i}$ as the rank of $y_{1i}$ among all $m + n$ observations. The Wilcoxon rank sum statistic for the case group is defined as $W = \sum_{i=1}^{m} R_{1i}$.

For simplicity, assume there are no ties for now. Note that $R_{1i}$ can be obtained by comparing $y_{1i}$ to each of the other $(m + n - 1)$ observations: $R_{1i} = 1 + \sum_{i'\neq i} I(y_{1i} > y_{1i'}) + \sum_{j=1}^{n} I(y_{1i} > y_{0j})$. After simplification, it is easy to see that the Wilcoxon rank sum statistic is the same as the Mann-Whitney U statistic, except for a constant term that depends only on the sample size:

$$W = m + \sum_{i=1}^{m} \sum_{i'=1, i'\neq i}^{m} I(y_{1i} > y_{1i'}) + \sum_{i=1}^{m}\sum_{j=1}^{n} I(y_{1i} > y_{0j}) = \frac{m(m+1)}{2} + U,$$

Here U is the Mann-Whitney U statistic, and it is defined as $U = \sum_{i=1}^{m}\sum_{j=1}^{n} I(y_{1i} > y_{0j})$. Thus, Wilcoxon rank-sum statistic W and Mann-Whitney's U statistic are mathematically equivalent. Furthermore, a large W or U indicates that $F_1$ is stochastically greater than $F_0$, while a small W or U indicates that $F_1$ is stochastically smaller than $F_0$ [35]. This implies that a test based on either statistic can be used to assess the null hypothesis $H_0$: $F_0 = F_1$.

*Probabilistic index (PI) to compare two subjects and the connection between PI and AUC*

When applied to one-sided, two-sample comparison problems, the rank-sum test is consistent for "stochastically greater" alternatives against the null hypothesis of the same underlying continuous distribution [35]. The rescaled rank-sum statistics (or equivalently U statistic) is a consistent estimator of $P(Y_1 > Y_0)$ or $P(Y_1 > Y_0) + \frac{1}{2}P(Y_1 = Y_0) > 0.5$ to accommodating tied pairs by assigning a weight of one half to a tie [36]. Because the value $P(Y_1 > Y_0) + \frac{1}{2}P(Y_1 = Y_0)$ quantifies the magnitude of "greater", it is referred to as the probabilistic index (PI) [37] or relative marginal effect [38]. In power analysis of the rank-sum test, PI is used as the effect size [37] and can be interpreted as the treatment effect. Now consider a random sample form $F_1$ and a random sample from $F_0$, denoted by $(y_{11}, \cdots, y_{1m})$ and $(y_{01}, \cdots, y_{0n})$, respectively. A consistent, unbiased, and minimum variance estimator of PI is the rescaled *U* statistic [36]:

$$\hat{PI} = \frac{1}{nm}\sum_{i=1}^{m}\sum_{j=1}^{n}\left[I(y_{1i} > y_{0j}) + \frac{1}{2}I(y_{1i} = y_{0j})\right] = \frac{1}{mn}U.$$

An interesting observation is that the PI is identical to AUC, that is, the area under the receiver operating characteristic curve (ROC) when assessing the degree of separation for two distributions [39–41]. As illustrated in Figure 2.1 Stage 1, for single-cell data, PI / AUC between a case subject and a control subject measures whether a biological measure of interest (gene expression, protein level, etc) of the cells from a case is "greater" than that of cells from a control. PI reaches its maximum value of 1 when two distributions are completely separable, while it equals 0.5 when two distributions overlap completely. This connection offers an adequate justification for the use of PI, which is a re-scaled rank sum/U statistic, to measure the difference between the expression level of a case to that of a control.

*Existing Robust Methods for Clustered Data*

While rank-based methods have gained widespread acceptance as robust alternatives to parametric approaches, there is a recognized need for extensions to address complex experimental designs and data structures. For instance, longitudinal outcomes often exhibit clustering within experimental units, and in observational studies, adjusting for covariates becomes essential when assessing the impact of a factor on a response variable that might also depend on other factors. In statistical methods, the incorporation of dependence structure in clustered data is handled in different ways. For instance, in [42], correlations of various types of pairwise comparisons were used to estimate the variance of a U statistic using pooled data for each condition. In another approach presented in [43], a new statistic is developed by taking the average of all possible rank-sum statistics from sampling one observation per cluster. While these methods prove suitable for data arising from randomized trials, extending them to accommodate covariate adjustments poses a considerable challenge.

Covariate adjustment is a standard practice in most statistical analyses, playing a crucial role in reducing bias due to confounders in observational studies. This is particularly important when the objective is to draw causal inferences. Even in randomized trials, including baseline covariates has been shown to increase statistical power [44]. Several methods have been proposed to address covariate adjustment in rank-based robust analysis. [45] introduced a conditional permutation test, while [46–49] developed weighted U statistics. Noteworthy works closely related to our proposed method include those by [50–52], which model pairwise comparisons using a logistic-type regression and employ a sandwich estimator for variance-covariance. In their more recent application [53], they analyze single-cell RNA-seq data by treating cells as independent observations. More recently, [54] introduced a U-statistic-based GEE method, [55] developed a sensitivity-based approach for adjusted Mann-Whitney test that provides sensitivity intervals using bootstrapped samples. Note that these robust methods mainly focus on covariate adjustment in studies with independent observations.

Our proposed method, ULV, extends these models in several significant ways. ULV is adaptable to clustered data, such as single-cell RNA-seq data. In ULV, we address correlations in pairwise comparisons through a latent variable model, which not only effectively characterizes dependence but also offers interpretable results. ULV allows for additional options such as covariate adjustment, comparisons between multiple groups, and weighted analysis for unbalanced cluster sizes.

### *Extended Wilcoxon rank-sum test for clustered data - a U-statistic-based Latent Variable (ULV) model*

Consider *m* cases and *n* controls, where each subject is associated with multiple measurements/cells. For the *i*th case, let $\vec{y}_{1i}$ represent a vector of length $K_{1i}$, i.e., $y_{1i} = (y_{1i1}, \cdots, y_{1iK_{1i}})^T$. The vectors of measurements for control subjects are defined similarly. We introduce the following U statistic

$$U_{cluster} = \sum_{i=1}^{m} \sum_{=1}^{n} h(\vec{y}_{1i}, \vec{y}_{0j}) \quad (1)$$

where $h(\vec{y}_{1i},\vec{y}_{0j})$ is a U statistic itself. Specifically, one robust choice is

$$h(\vec{y}_{1i},\vec{y}_{0j}) = \frac{1}{K_{1i}K_{0j}} \sum_{k=1}^{K_{1i}} \sum_{k'=1}^{K_{0j}} \left[ I(y_{1ik} > y_{0jk'}) + \frac{1}{2} I(y_{1ik} = y_{0jk'}) \right] .$$

It is not difficult to see that this choice of $h(\vec{y}_{1i},\vec{y}_{0j})$ is an estimate of the probability index (PI) when comparing $\vec{y}_{1i}$ and $\vec{y}_{0j}$. As a rank-based statistic, it offers robustness against distributional assumptions and the presence of outliers.

One way to obtain p-values is to employ a permutation method, which generates statistics under the null hypothesis by shuffling the labels of cases and controls. However, this method can be computationally intensive, especially when thousands or tens of thousands of tests must be performed, as in gene expression analysis. Additionally, it tends to have low power in scenarios with small values of $m$ and $n$. The asymptotic distribution of the $U_{cluster}$ statistic can be derived by obtaining an estimator of its variance using Hijek projection [56]. Instead, we propose a new framework aimed at achieving broader objectives. Specifically, to overcome the limitations of traditional approaches, we propose to model the pairwise differences or monotonically transformed data using latent variables, as shown below.

### *A latent variable model for two-group comparisons*

Here we elaborate on how to model the pairwise differences using a latent variable approach. In this framework, we assume that the expression level of a subject is unobserved, and the resulting model for the pairwise differences is equivalent to a two-way model with a single observation per combination. From an experimental design perspective, this approach allows us to better understand and characterize individuals as independent entities, as opposed to the observational units (e.g., cells), which might be highly correlated among those from the same subject. Let

$$d_{ij} = h(\vec{y}_{1i}, \vec{y}_{0j}), i = 1,\cdots,m; j = 1,\cdots,n \qquad (2)$$

denote the difference between the observations from the $i$th case and the $j$th control.

We model $d_{ij}$ as

$$d_{ij} = a_i - b_j + \epsilon_{ij}, \ i = 1,\cdots,m; j = 1,\cdots,n \qquad (3)$$

where $a_i$ represents the latent (relative and not directly observed) expression level of the $i$th case, $b_j$ is the latent level of the $j$th control. We assume that $a_i$'s, $b_j$'s, and $\epsilon_{ij}$'s are mutually independent random variables following normal distributions:

$$a_i \sim N(\mu, \sigma_1^2), b_j \sim N(0, \sigma_0^2), \epsilon_{ij} \sim N(0, \sigma^2).$$

In this model, $\mu$ is the difference in the population levels between the case group and the control group, as the population mean of the control group is fixed at 0. Inference of $\mu$ can be obtained by fitting a linear mixed-effects model. The null hypothesis of no differential expression can be rewritten as $H_0$: $\mu$ = 0.5 when the function $h()$ is the estimated probability index between two subjects or as $H_0$ : $\mu$ = 0 when the logit transformation is instead used.

We offer some remarks: (1) we describe the $a_i$'s and $b_j$'s as the relative but not absolute levels because adding any arbitrary constant to both $a_i$ and $b_j$ does not change their difference. (2) Since the mean of the $b_j$'s is assumed to be zero, there is no difference between adding or subtracting $b_j$ from $a_i$; however, using the difference rather than the sum emphasizes that the quantity modeled is the pairwise differences. (3) The latent model is also a two-way random-effects model with a single observation per cell; in this case, a closed-form test statistic exists, as we will show later. (4) Finally, the latent-level model we adopted has been widely used in psychometric fields. One example is the Rasch model [57], which uses a logit link to model the probability of answering a question correctly by associating the binary correct-incorrect outcome with a person-specific ability parameter and a question-specific difficulty parameter.

### *Adjusting for covariates*

The latent variable approach we proposed is a regression framework, which can model effects from covariates naturally. Suppose there are $p$ covariates, denoted by $x_i = (x_{i1}, \cdots x_{ip})^T$ for the $i$th subject, and analogously for the $j$-th subject. Then the pairwise differences can be modeled as

$$\begin{aligned} d_{ij} &= (a_{0i} + \beta^T x_i) - (b_{0j} + \beta^T x_j) + \epsilon_{ij} \\ &= a_{0i} - b_{0j} + \beta^T (x_i - x_j) + \epsilon_{ij} \end{aligned} \tag{4}$$

where $a_{0i}$ and $b_{0j}$ represent the covariate-adjusted latent levels. As in the model without covariates, the random effects in Model 4 are mutually independent with normal distributions:

$$a_{0i} \sim N(\mu, \sigma_1^2), b_{0j} \sim N(0, \sigma_0^2), \epsilon_{ij} \sim N(0, \sigma^2)$$

Again, this is a linear mixed-effects model where $\mu$ and $\beta$ denote fixed-effects coefficients for the parameter of interest and the parameters related to covariates, respectively.

*Multi-group comparisons*

We can extend the two-condition comparison framework to accommodate scenarios involving more than two conditions. Consider a scenario with $M$ + 1 total conditions, we can select one condition as the reference condition and compare all the remaining conditions to it. The pairwise difference between the $i$th subject under condition $m$ and the $j$th subject in the reference condition is modeled as:

$$d_{ij}^{m} = a_{mi} - b_{0j} + \epsilon_{ij}, \quad i = 1, \cdots, N_m; \quad j = 1, \cdots, N_0; \quad m = 1, \cdots, M, \qquad (5)$$

where $a_{mi}$ and $b_{0j}$ denote, respectively, the latent levels of the $i$th subject in condition $m$ and the $j$th subject in the reference condition. Here, $N_0$ and $N_m$ represent the number of subjects in the reference condition and in condition $m$. The parameter $\mu_m$ represents the mean difference in latent levels between condition $m$ and the reference condition, defined as

$$a_{mi} \sim N(\mu_m, \sigma_m^2), \quad b_{0j} \sim N(0, \sigma_0^2), \quad \epsilon_{ij} \sim N(0, \sigma_\epsilon^2) \qquad (6)$$

The null hypothesis of no difference can be expressed as $H_{0:}\ \mu_1 = \cdots = \mu_M = \mu_0$ and can be tested using a Wald or a likelihood ratio test for linear mixed-effects models.

*A weighted analysis for varying cluster sizes*

The final issue we will address is varying cluster sizes, which is common in many clustered datasets, such as single-cell data. For instance, while one subject may have a few thousand cells measured, another might have an order of magnitude more or less measured. This variation necessitates a reconsideration of the standard assumption of equal variance for latent variables. As discussed previously, the probabilistic index (PI) between the $i$th case and the $j$th control is a rescaled U statistic. Following the variance formula of the Mann-Whitney U statistic, the variance of PI is a function of the cluster sizes; specifically, it is proportional to ($1/K_{0i} + 1/K_{1j}$), where $K_{1i}$ and $K_{0j}$ denote the cluster sizes of the $i$th case and $j$th control, respectively. For this reason, we adapt our model by relaxing the assumption of equal variance for latent variables under the same treatment. Specifically, we propose that their variances should be inversely proportional to their respective cluster sizes:

$$a_i \sim N(\mu, \frac{\sigma_1^2}{K_{1i}}), b_j \sim N(0, \frac{\sigma_0^2}{K_{0j}})$$

For hypothesis testing in this modified framework, both the Wald test and the likelihood ratio test are suitable options.

### *Background of single-cell RNA-seq data and existing methods*

The methodology work we proposed is motivated by the challenges in analyzing single-cell data, especially single-cell RNA-seq (scRNA-seq), a powerful biological method for measuring the transcriptomic activities of individual cells. Compared to earlier techniques in measuring gene expression levels, scRNA-seq provides a more detailed, comprehensive, and accurate view of gene expression patterns that may be missed by traditional bulk RNA-seq techniques. Since its inception, scRNA-seq has become an important tool in many areas of biological and medical research. For example, it has also been used to identify and characterize new cell types, to study highly heterogeneous cell populations such as those found in cancer or stem cells, and to investigate the role of gene expression in diseases.

A fundamental goal in gene expression studies is to identify genes that show differential expression across different groups or conditions. The DESeq2 method [9], originally designed for analyzing bulk sequencing data, models read counts using negative binomial distributions. However, in the widespread application of DESeq2 to scRNA-seq data, investigators have treated cells, rather than subjects, as independent units, as illustrated in the right most plot of Figure 1). The misuse of bulk sequencing methods, such as DESeq2 and other popular methods edgeR [8] and MAST [10], has led to unacceptably high false positive rates. Pseudobulk approaches (sketched in the left most plot of Figure 1), which combine cells within subjects before applying DESeq2 or similar methods to bulk sequencing data, have been recommended to reduce false positives caused by ignoring the cluster structure in scRNA-seq data.

### *Simulating gene counts based on parameters estimated from real data*

To ensure the validity of our simulation study, it is essential to choose a simulation method that can generate data resembling real scRNA-seq data. To achieve this, we adopted the simulation strategy used in the BSDE method [7], which generates gene count data from a zero-inflated negative binomial model with parameters estimated from published real scRNA-seq data. Specifically, using a reference dataset, we estimated the mean parameter $\mu_{ij}$, the dispersion parameter $\phi_{ij}$, the dropout rate parameter $z_{ij}$, and the within-subject variability $\sigma_{ij}$ of expression levels for gene $i$ and individual $j$. For cell $k$, the gene expression level $Y_{ijk}$ of gene $i$ and individual $j$ is simulated using zero-inflated negative binomial (ZINB) distribution as follows,

$$\mu_{ijk} \sim N(\mu_{ij}, \sigma_{ij}^2), \quad (7)$$

$$Y_{ijk} \sim ZINB(\mu_{ijk}, \phi_{ij}, z_{ij}). \quad (8)$$

The reference dataset mentioned in [7] consists of the single cell gene expression matrix of 100 genes from multiple subjects. To simulate data under the null hypothesis, we first resample the estimated parameters from each person-gene combination until the desired number of genes and subjects are obtained. Then, we generate read counts from zero-inflated negative binomial distributions.

To simulate data under the alternative hypotheses, we introduce the fold change parameter *r*, with *r* = 1 representing non-DE genes and *r* > 1 indicating overexpressed genes. To do this, we modified the mean and dispersion parameters for the case group in accordance with the Mean DE simulation method outlined in [7]. Specifically, relative to the mean parameter $\mu$ and the dispersion parameter $\phi$, a fold change of *r*, is achieved if the mean and dispersion parameters $\mu'$ and $\phi'$ are defined as follows,

$$\mu' = \frac{\mu}{r}, \phi' = \frac{\phi\mu}{\mu + (1-r)\phi}. \quad (9)$$

In simulations involving covariates, we consider the following framework:

$$log\left(\mu_{ij}^*\right) = log(\mu_{ij}) + \beta x_j \quad (10)$$

where $\mu_{ij}^*$ represents the subject-level mean for subject *j*, $x_j$ represents the covariate value of the subject *i* and $\beta$ denotes the corresponding regression coefficient. Here, a $\beta$ value of 0 indicates no effect on the outcome while 0.5 represents an effect of the covariate on the outcome. To create different distributions of the covariate variable for a two-group study, we assign the covariate variable values equally spaced between -0.9 and 1.1 for the case group and values between -1 to 1 for the control group. Note that the covariate variable is fixed rather than being generated from a distribution due to the small sample size (5 or 10 subjects in each group) considered in our study.

In addition to the simulation methods described above, we also perform additional simulation studies to assess the performance of our ULV method and six benchmarking methods. The six methods include DESeq2 (both the pseudobulk and single-cell versions), MAST (both the original version without subject-specific effects and the revised version with subject specific effects), and the Wilcoxon-Mann-Whitney test (applied both at the cell level and at the subject level). For each simulation scenario, data will be generated 100 times, allowing us to explore and illustrate the variance in the results obtained.

*Additional simulation parameters to assess the Type I error rates*

In addition to the simulation study described in the previous section, we have also evaluated type I error rates of our method and those of competing methods under multiple scenarios that vary with respect to the number of cells per subject, the significance level, and the sample size considered. Specifically, cell numbers were varied from 20 to 400 to represent relatively rare to common cell types. We considered 4 significance levels: from 0.2 representing a very liberal cutoff, 0.05 denoting the typical cutoff used in the absence of multiple comparisons, to 0.01 and 0.001. We present results for simulations that comprise 5 or 10 subjects per group.

*Simulations with unequal cluster sizes*

The simulation method described includes a common and fixed number of cells per subject. However, in real studies, an additional challenge is the unequal (unbalanced) cluster sizes. To accommodate this, we have introduced weighted ULV. To assess the performance of weighted ULV, we conducted simulation studies with unequal cluster sizes, obtained by letting the number of observations per individual vary according to a discrete uniform distribution between 50 and $N_{max}$, where $N_{max} = 200$ or 400. As in previous analyses, a fold change $r = 1$ is used for estimating type I error rates while $r = 1.5$ or 2 for determining statistical power. The number of subjects per condition was kept either 5 or 10 to understand the effect of sample size on the results.

*Generation of subject label-permuted datasets for type I error rate assessment*

To assess the type I error rate of various DE methods, we permuted the subject-level labels between two conditions from the original single-cell dataset to generate null data. Specifically, for the pediatric AML single-cell proteomics dataset, we conducted the permutation analysis on the cell type of HSPC from 5 healthy controls and 16 AML patients. To ensure balanced groups in the permuted dataset, we excluded 1 healthy control subject due to the age and gender discrepancy, and next we placed 8 AML and 2 healthy control subjects in one group, and placed the rest in another group. From a total of 38610 possible permutation sets, we randomly selected a sample of 100 sets to calculate the type I error rate for three different DE methods (ULV_wt, Wilcoxon_sc, and Wilcoxon_pseudobulk). This permutation approach ensures a robust evaluation of the type I error rates across DE methods.

*Connection between ULV and the two-way random-effects model*

As discussed in Section 4.4.1, there is no mathematical difference between using $a_i - b_j$ or $a_i + b_j$ and our choice of the former reflects the fact that pairwise comparisons $d_{ij}$'s are differences. It is clear that the latent variable model in Equation 3 is a two-way random-effects model with a single observation per cell. It is easy to verify that an unbiased estimator of $\mu$ is $\hat{u} = \bar{d}_{..}$; in fact, it can also be shown that it is the MLE. Its variance is

$$var(\hat{\mu}) = \frac{\sigma^2}{mn} + \frac{\sigma_1^2}{m} + \frac{\sigma_0^2}{n},$$

which can be estimated using sums of squares, as given below

$$v\hat{a}r(\hat{\mu}) = \frac{1}{m}\hat{\sigma}_1^2 + \frac{1}{n}\hat{\sigma}_0^2 + \frac{1}{mn}\hat{\sigma}^2 \quad , \qquad (11)$$

where

$$\hat{\sigma}_0^2 = \frac{\sum_{i=1}^{m}(\bar{d}_{i\cdot} - \bar{d}_{..})^2}{m-1}, \hat{\sigma}_1^2 = \frac{\sum_{j=1}^{n}(\bar{d}_{\cdot j} - \bar{d}_{..})^2}{n-1}, \hat{\sigma}^2 = \frac{1}{(m-1)(n-1)}\sum_{i=1}^{m}\sum_{j=1}^{n}(d_{ij} - \bar{d}_{i\cdot} - \bar{d}_{\cdot j} + \bar{d}_{..})^2 \quad .$$

In practice, we found that the last term of Equation 11 is small compared to the other terms, indicating that

$$v\hat{a}r(\hat{\mu}) \approx \frac{1}{m}\hat{\sigma}_1^2 + \frac{1}{n}\hat{\sigma}_0^2 \quad , \qquad (12)$$

Recall that the null hypothesis of no difference between the case and control populations is $H_0 : \mu = \mu_0$ with $\mu_0 = 0.5$ of probabilistic index and $\mu_0 = 0$ for its logit transformation. The following t-test

$$t = \frac{\bar{d}_{..} - 0.5}{\sqrt{\frac{1}{m}\hat{\sigma}_1^2 + \frac{1}{n}\hat{\sigma}_0^2}} \quad , \qquad (13)$$

which follows $t_{m+n-2}$ or the standard normal distribution asymptotically when the null hypothesis is true.

*An analytical solution and least squares estimate*

The closed-form t-test in Equation 13 greatly reduces the computational cost when there is no covariate to be adjusted for. One interesting observation is that the t-statistic is identical to the two-sample t-statistic based on $(\hat{a}_1, . . . \hat{a}_m)$ and $(\hat{b}_1, . . . \hat{b}_n)$, where

$$\hat{a}_i = \bar{d}_{i\cdot}, \hat{b}_j = \bar{d}_{\cdot j} - \bar{d}_{..}, i = 1, \cdots, m; j = 1, \cdots, n, \qquad (14)$$

which is one of the infinitely many least squares solution to the problem in Equation 3. In the following, we will show that least squares solutions exist and are unique up to a constant; this makes sense intuitively as adding a common constant to the latent levels does not affect the differences.

**Theorem**. Consider the model $d_{ij} = a_i - b_j + \epsilon_{ij}$. Then least squares solutions for $a_i$ and $b_i$ are unique up to a constant.

**Proof**. Let $d = (d_{11}, \cdots, d_{1n}, \cdots, d_{m1}, \cdots, d_{mn})^T$ denote the vector of pairwise of differences, $a = (a_1 \cdots, a_m)^T$ and $b = (b_1, \cdots, b_n)^T$ denote the vector of latent levels for the case group and control group, respectively. Further, the design matrix is

$$\mathbf{D} = \begin{pmatrix} \mathbf{1}_n & \cdots & \cdots & \mathbf{0}_n & -\mathbf{I}_n \\ \mathbf{0}_n & \mathbf{1}_n & \cdots & \mathbf{0}_n & -\mathbf{I}_n \\ \vdots & & \ddots & & \vdots \\ \mathbf{0}_n & \cdots & \cdots & \mathbf{1}_n & -\mathbf{I}_n \end{pmatrix}_{(mn)\times(m+n)}$$

It is not difficult to verify that *rank*(**D**) = $m + n - 1$ and every LSE solution can be expressed as a generalized inverse of **D** multiplied by $d$. Of particular interest is the solution in Equation 1, which is obtained by applying the linear constraint $\sum_{j=1}^{n} b_j = 0$. Note that this linear constraint is consistent with our ULV, which assumes that the $b_j$'s are from a normal distribution centered at zero.

Another interesting solution is when using the Moore-Penrose inverse of **D**. This leads to:

$$\begin{pmatrix} \hat{a} \\ \hat{b} \end{pmatrix} = \mathbf{D}^{+}\mathbf{d} = \begin{pmatrix} \bar{d}_{1\cdot} + \frac{m}{m+n}\bar{d}_{\cdot\cdot} \\ \vdots \\ \bar{d}_{m\cdot} + \frac{m}{m+n}\bar{d}_{\cdot\cdot} \\ \bar{d}_{\cdot 1} - \frac{n}{m+n}\bar{d}_{\cdot\cdot} \\ \vdots \\ \bar{d}_{\cdot n} - \frac{n}{m+n}\bar{d}_{\cdot\cdot} \end{pmatrix},$$

where $\mathbf{D}^{+}$ is the Moore-Penrose generalized inverse, which is the generalized inverse with the minimum Frobenius norm among all generalized inverses of **D**.

A final remark is that the LSE are exact solutions for separable difference metrics, i.e., the difference function between case $i$ and control $j$ can be expressed as

$$d_{ij} = h(\vec{y}_{1i}, \vec{y}_{0j}) = g(\vec{y}_{1i}) - g(\vec{y}_{0j})$$

for some function $g$.

**Proof.** When this function $h(;)$ can be separated, i.e. $h(\vec{y}_{1i}, \vec{y}_{0j}) = g(\vec{y}_{1i}) - g(\vec{y}_{0j})$, it is not difficult to verify that both the coefficient matrix **D** and the augmented matrix [**D**⁝**d**] have rank $m + n - 1$. By the Rouch´e–Capelli theorem, the set of equations

$$D \begin{pmatrix} a \\ b \end{pmatrix} = d$$

is consistent. As a result, the solutions are exact and one solution is $\hat{a}_i = g(\vec{y}_{1i})$, $\hat{b}_j = g(\vec{y}_{0j})$, for $i=1...m$, $j = 1...n$.

Separable difference metrics include many pseudobulk methods that compare subject-summary data between groups. One such example is $g(\vec{x}_{1i}) = \bar{x}_{1\cdot}$, in which case the mean of a subject is used in place of the observations for all the cells and the resulting test is a two-sample t-test based on subject means. From this point of review, our framework covers well known situations when the pairwise difference is defined in specific ways.

### *Data Availability*

The single-cell proteomics data from AML patients and healthy controls, which was generated by Lasry et al. [12], is available from Single-Cell Proteomic Database (SCPD)[58] at https://scproteomicsdb.com/data info/Dataset93/None. The scRNA-seq data from COVID-19 patients, which was generated by Schulte-Schrepping et al. [19], is available from https://beta.fastgenomics.org/home by searching key word *Schulte-Schrepping* in the data section.

### *Code Availability*

ULV is available as an R package at GitHub (https://github.com/yu-zhaoxia/ULV).

Legends

**Fig. 1 An overview of two common ways of analyzing single-cell data, along with our proposed method.** Left: pseudo-bulk analysis, which first aggregates the measurements of all cells within a biological unit and then conducts differential analysis across experimental conditions; Right: cell-level analysis, which treats cells as independent observational units; Middle: our proposed method, which models pairwise differences using latent variables, serving as an innovative compromise and hybrid of the other two methods.

**Fig. 2 ULV combines pairwise single-cell comparisons with sample wide approaches to identify differentially expressed genes. a**: Data preprocessing, such as quality control and normalization, for production of single-cell level data with high quality. The single-cell RNA-seq levels of a gene from 5 controls and 10 cases are presented using box plots, with each box plot visualizing the expression level of all the cells of a subject. **b**: A matrix of pairwise differences is calculated, with the (i,j)th entry denoting the difference between the *i*th case and the *j*th control. Our recommended difference metric is the probability index (PI) because it is a robust measure of "greater than" of one subject to another; additionally, PI is closely connected to Wilcoxon rank sum U statistics and is equivalent to the area under the receiver operating characteristic curve for binary classification; **c**: The matrix of pairwise PIs from **b** is modeled using a latent variable model. This linear model framework allows adjusting for covariates, generalizing to multiple groups and weighted analysis, and applying data transformation such as the logit transformation of the PI.

**Fig. 3 ULV maintains lower type I error rates and higher power when compared with pseudobulk differential expression methods.** The type I error rate (**a-c**) and statistical power (**d**, **e**) of ULV and six competing DE methods are summarized. **a** Type I error rate of ULV and six competing DE methods, indicated by different colors. The horizontal axis displays the cell numbers per subject. Each simulated dataset included 5 cases and 5 controls. The horizontal dashed lines represent significance levels. The results for the three single cell-based methods are not fully shown because their values fall outside of the range of the figures. Each boxplot is based on 100 simulations using the same model parameters. **b** Type I error rates of ULV and six competing DE methods for covariate adjustment. Covariates are not included in the two Wilcoxon methods because these methods do not allow for covariate adjustment. Each simulated dataset includes 5 cases and 5 controls, with 100 cells measured from each subject. The covariate coefficient was set as 0 (Left) and 0.5 (Right). The horizontal dashed line indicates the significance levels at $\alpha = 0.05$. ULV_adj refers to the ULV method adjusted for covariates, while ULV represents the ULV method without covariate adjustment. Each box plot is based on 100 simulations using the same model parameters. **c** Similar to **b,** but with no simulated covariates. To mimic varying cluster sizes in real studies, cluster sizes were simulated from a uniform distribution of (50, $N_{max}$) where $N_{max} = 200$ or 400. The dashed lines represent the desired significance level. Each box plot is based on 100 simulations using the same model parameters. ULV_wt denotes ULV corrected for disparate cluster sizes. **d** Statistical power of ULV, ULV_wt, and six competing methods when fold change $r = 1.5$. The other simulation details follow those described in (**c**). **e** Statistical power of ULV, ULV_wt, and six competing methods when fold change $r = 2$. The other simulation details follow those described in (**c**).

**Fig. 4 ULV identifies differentially regulated proteins missed by Wilcoxon pseudobulk tests. a**, HSPC and myeloid cell counts from pediatric AML patients and age-matched healthy controls. **b**, Volcano plots of cell type-specific differential protein expression analysis via ULV_wt. The significantly differential proteins in HSPC and myeloid cells are highlighted in red. **c**, Representative examples of differentially expressed proteins in HSPC and myeloid cells between AML patients and healthy controls. The raw expression values were pre-processed via centered log ratio normalization. CD123: Interleukin 3 receptor alpha. CD279-PD-1: Programmed cell death protein 1. **d**, Type I error rate estimates of different DE methods when p-value cutoff is 0.005, 0.01 and 0.05 using label-permuted dataset of HSPC cells.

**Fig. 5 Inclusion of covariates is critical for single-cell RNA differential expression. a**, CD8+ T cell counts in 5 mild and 10 severe COVID subjects and the subject level covariate information including age and gender. **b**, A gene displaying gender-dependent expression. In the heatmap, the rows and columns represent severe and mild COVID subjects respectively and are ordered by gender. The scatter plot provides the severe-mild difference in terms of probabilistic index by different comparison groups based on gender. The volcano plot highlights (in red) the significant DE genes after adjusting for gender. Here a gene was considered

differentially expressed if its FDR-adjusted p-value is below 0.1 and its probabilistic index is below 0.45 or above 0.55. The top 15 genes with the largest difference compared with unadjusted ULV in terms of the estimated probabilistic index or FDR-adjusted p-values are marked. **c**, A gene displaying age-dependent expression. In both heatmaps, the rows and columns are ordered by age. The scatter plot showing the positive correlation between severe-mild pairs in age difference and probabilistic index. The volcano plot highlights (in red) the significant DE genes after adjusting for age. The top 15 genes with the largest difference compared with unadjusted ULV in terms of the estimated probabilistic index or FDR-adjusted p-values are marked. **d**, Venn diagram demonstrating the number of differentially expressed genes identified using ULV, ULV_adj (both covariates), and by both methods.

**Fig. 6. ULV exhibits decreased expression bias and uniquely identifies critical COVID-19 related pathways when compared with other differential expression bias.**
**a-b**, Box plots of mean expression level (**a**) and proportions of zero expression measurements (**b**) of the 100 top-ranked genes from each DE method. **c**, Increase in enrichment for relevant gene sets through GSEA using the ranked genes by ULV_wt and DESeq2_pseudobulk respectively, after adjustment for age and gender. The Venn diagram shows the number of pathways identified by each of the two methods, or both methods, with the top 5 pathways listed for each group. **d**, The significance levels of the 10 top pathways identified by ULV_wt but not by DESeq2_pseudobulk are provided as a bar plot.

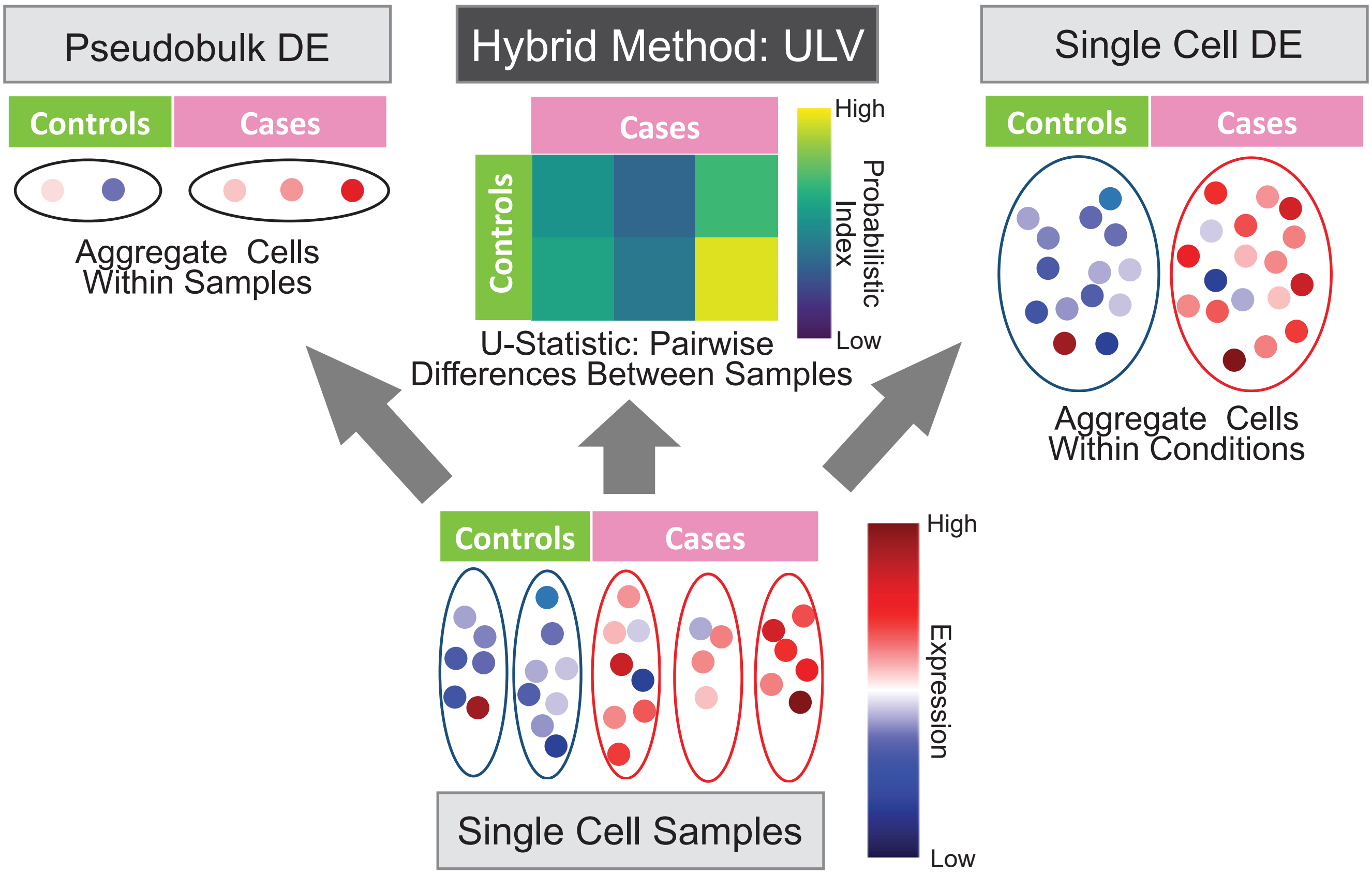

Fig 1.

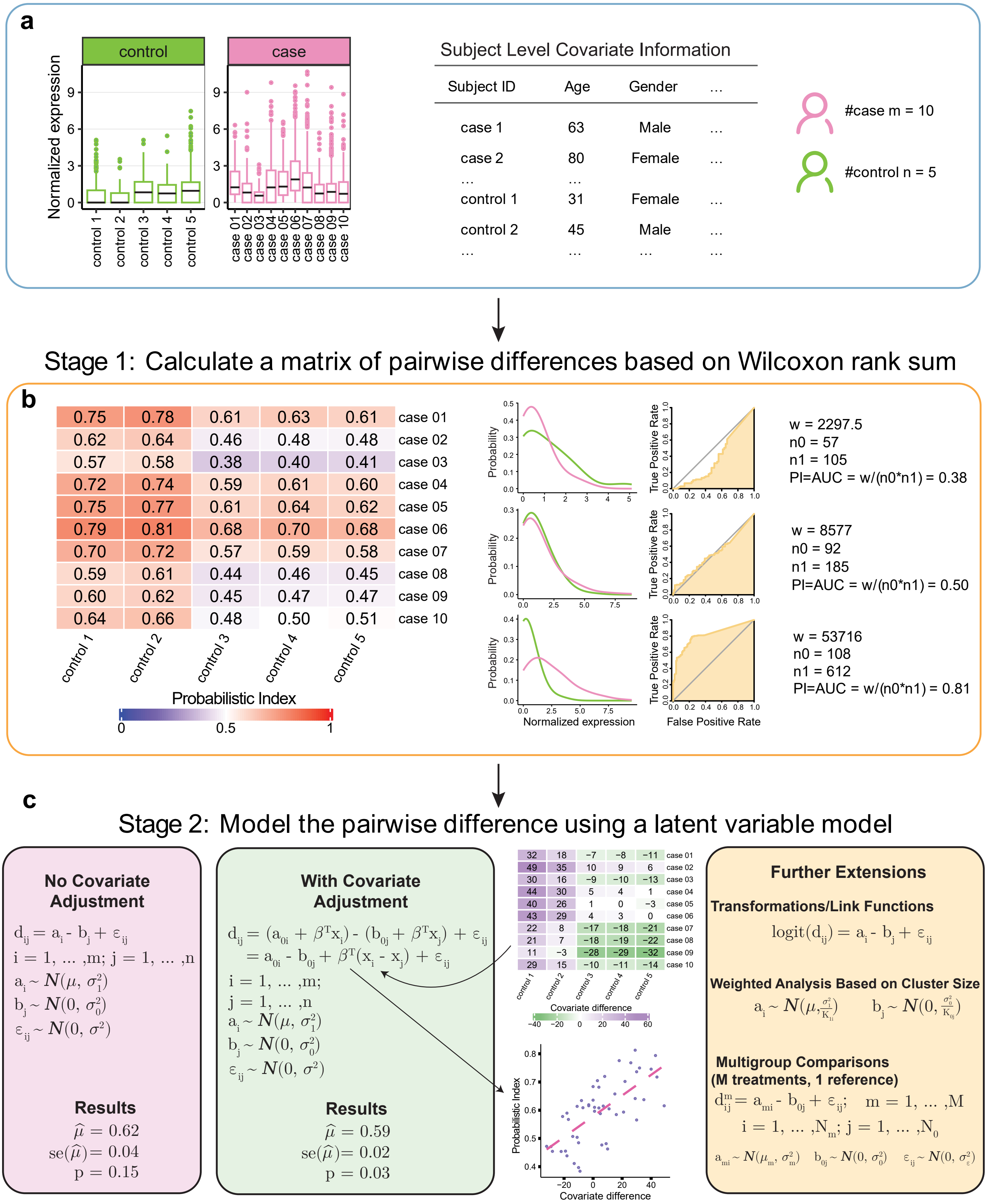
a
control
case
Normalized expression
Subject Level Covariate Information
Subject ID Age Gender ...
case 1 63 Male ...
case 2 80 Female ...
control 1 31 Female ...
control 2 45 Male ...
#case m = 10
#control n = 5
Stage 1: Calculate a matrix of pairwise differences based on Wilcoxon rank sum
b
Probabilistic Index
Probability
True Positive Rate
Normalized expression
False Positive Rate
w = 2297.5
n0 = 57
n1 = 105
PI=AUC = w/(n0*n1) = 0.38
w = 8577
n0 = 92
n1 = 185
PI=AUC = w/(n0*n1) = 0.50
w = 53716
n0 = 108
n1 = 612
PI=AUC = w/(n0*n1) = 0.81
c
Stage 2: Model the pairwise difference using a latent variable model
No Covariate Adjustment
With Covariate Adjustment
Results
Covariate difference
Probabilistic Index
Further Extensions
Transformations/Link Functions
Weighted Analysis Based on Cluster Size
Multigroup Comparisons (M treatments, 1 reference)

Fig 2.

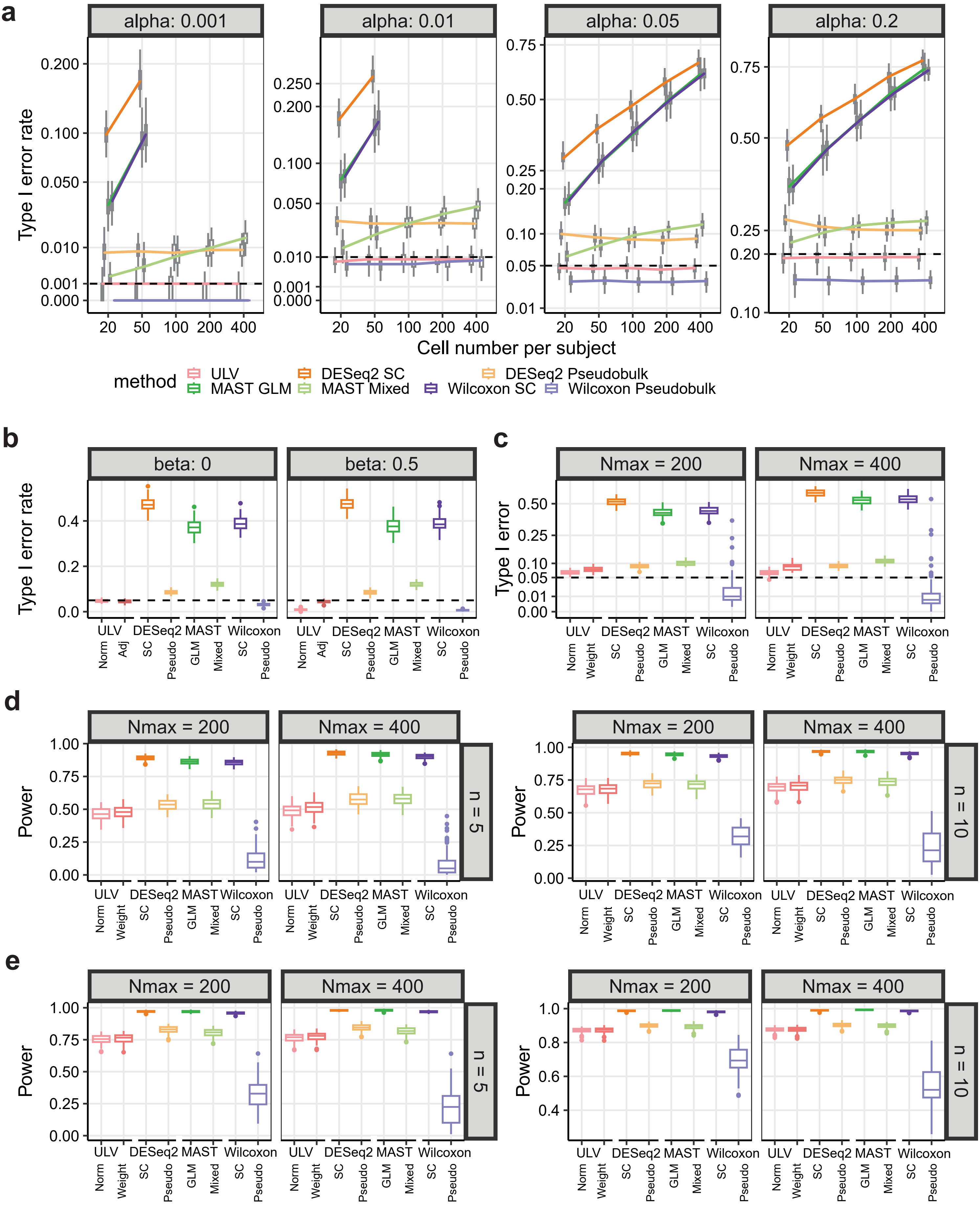

Fig 3.

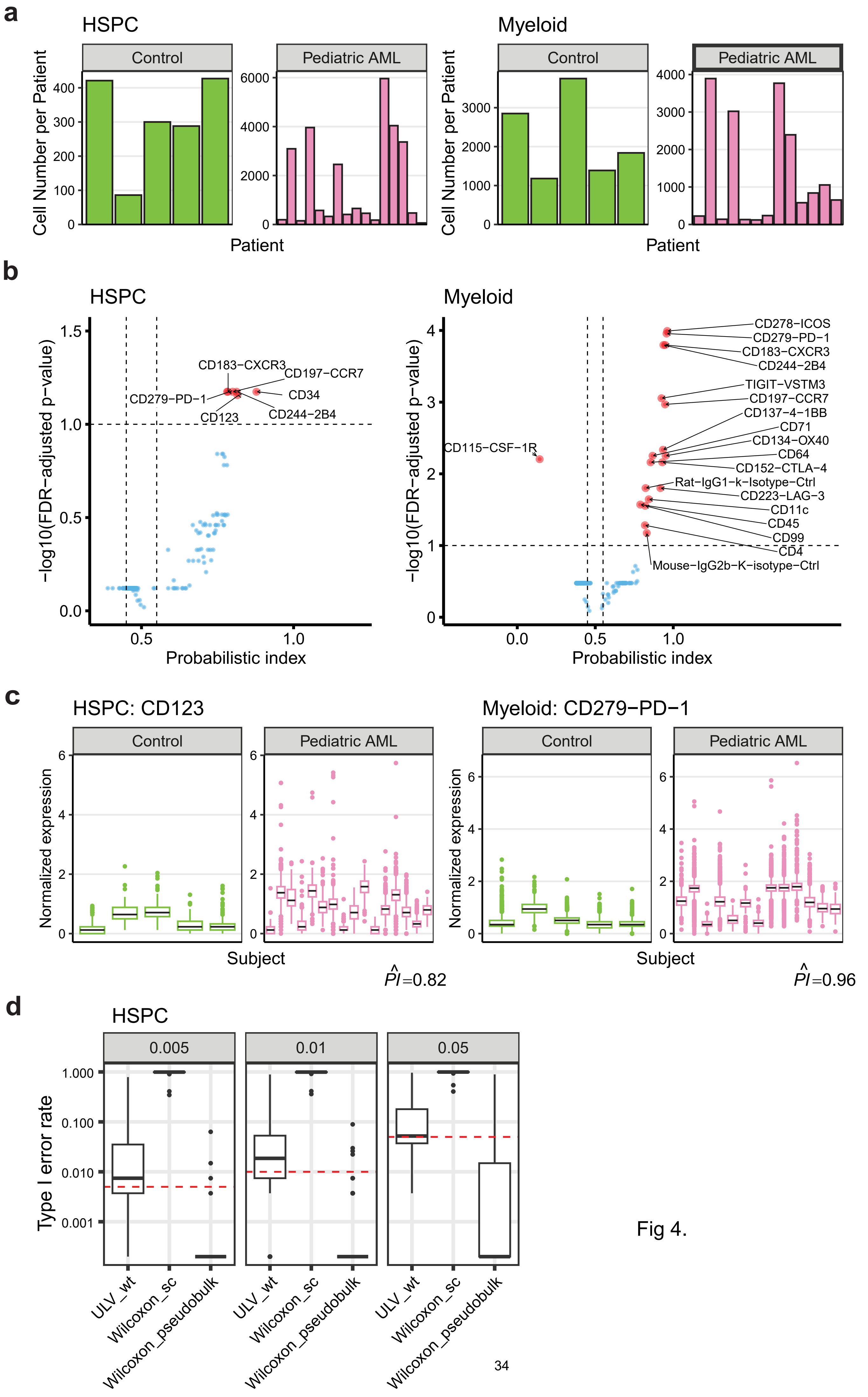

Fig 4.

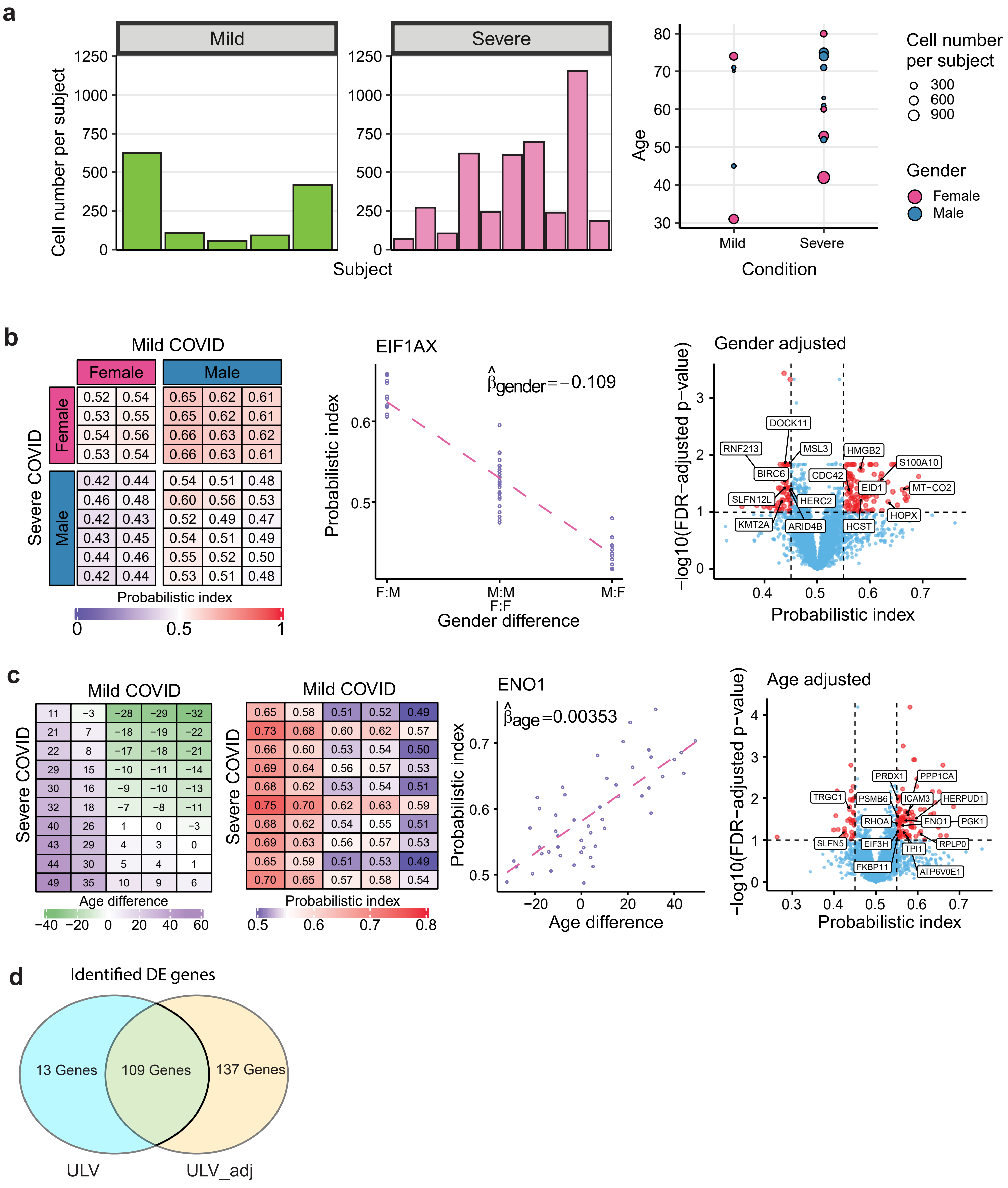

a
Mild
Severe
Cell number per subject
1250
1000
750
500
250
0
Subject
Age
80
70
60
50
40
30
Mild
Severe
Condition
Cell number per subject
300
600
900
Gender
Female
Male
b
Mild COVID
Female
Male
Severe COVID
Female
Male
0.52 0.54 0.65 0.62 0.61
0.53 0.55 0.65 0.62 0.61
0.54 0.56 0.66 0.63 0.62
0.53 0.54 0.66 0.63 0.61
0.42 0.44 0.54 0.51 0.48
0.46 0.48 0.60 0.56 0.53
0.42 0.43 0.52 0.49 0.47
0.43 0.45 0.54 0.51 0.49
0.44 0.46 0.55 0.52 0.50
0.42 0.44 0.53 0.51 0.48
Probabilistic index
0
0.5
1
EIF1AX
$\hat{\beta}_{gender}=-0.109$
Probabilistic index
0.6
0.5
F:M
M:M
F:F
M:F
Gender difference
Gender adjusted
−log10(FDR−adjusted p−value)
3
2
1
0
DOCK11
RNF213
MSL3
HMGB2
S100A10
BIRC6
CDC42
EID1
MT−CO2
SLFN12L
HERC2
HOPX
KMT2A
ARID4B
HCST
0.4
0.5
0.6
0.7
Probabilistic index
c
Mild COVID
Severe COVID
11 −3 −28 −29 −32
21 7 −18 −19 −22
22 8 −17 −18 −21
29 15 −10 −11 −14
30 16 −9 −10 −13
32 18 −7 −8 −11
40 26 1 0 −3
43 29 4 3 0
44 30 5 4 1
49 35 10 9 6
Age difference
−40 −20 0 20 40 60
Mild COVID
Severe COVID
0.65 0.58 0.51 0.52 0.49
0.73 0.68 0.60 0.62 0.57
0.66 0.60 0.53 0.54 0.50
0.69 0.64 0.56 0.57 0.53
0.68 0.62 0.53 0.54 0.51
0.75 0.70 0.62 0.63 0.59
0.68 0.62 0.54 0.55 0.51
0.69 0.63 0.56 0.57 0.53
0.65 0.59 0.51 0.53 0.49
0.70 0.65 0.57 0.58 0.54
Probabilistic index
0.5 0.6 0.7 0.8
ENO1
$\hat{\beta}_{age}=0.00353$
Probabilistic index
0.7
0.6
0.5
−20
0
20
40
Age difference
Age adjusted
−log10(FDR−adjusted p−value)
4
3
2
1
0
PRDX1
PPP1CA
TRGC1
PSMB6
ICAM3
HERPUD1
RHOA
ENO1
PGK1
SLFN5
EIF3H
TPI1
RPLP0
FKBP11
ATP6V0E1
0.3
0.4
0.5
0.6
0.7
Probabilistic index
d
Identified DE genes
13 Genes
109 Genes
137 Genes
ULV
ULV_adj

Fig 5.

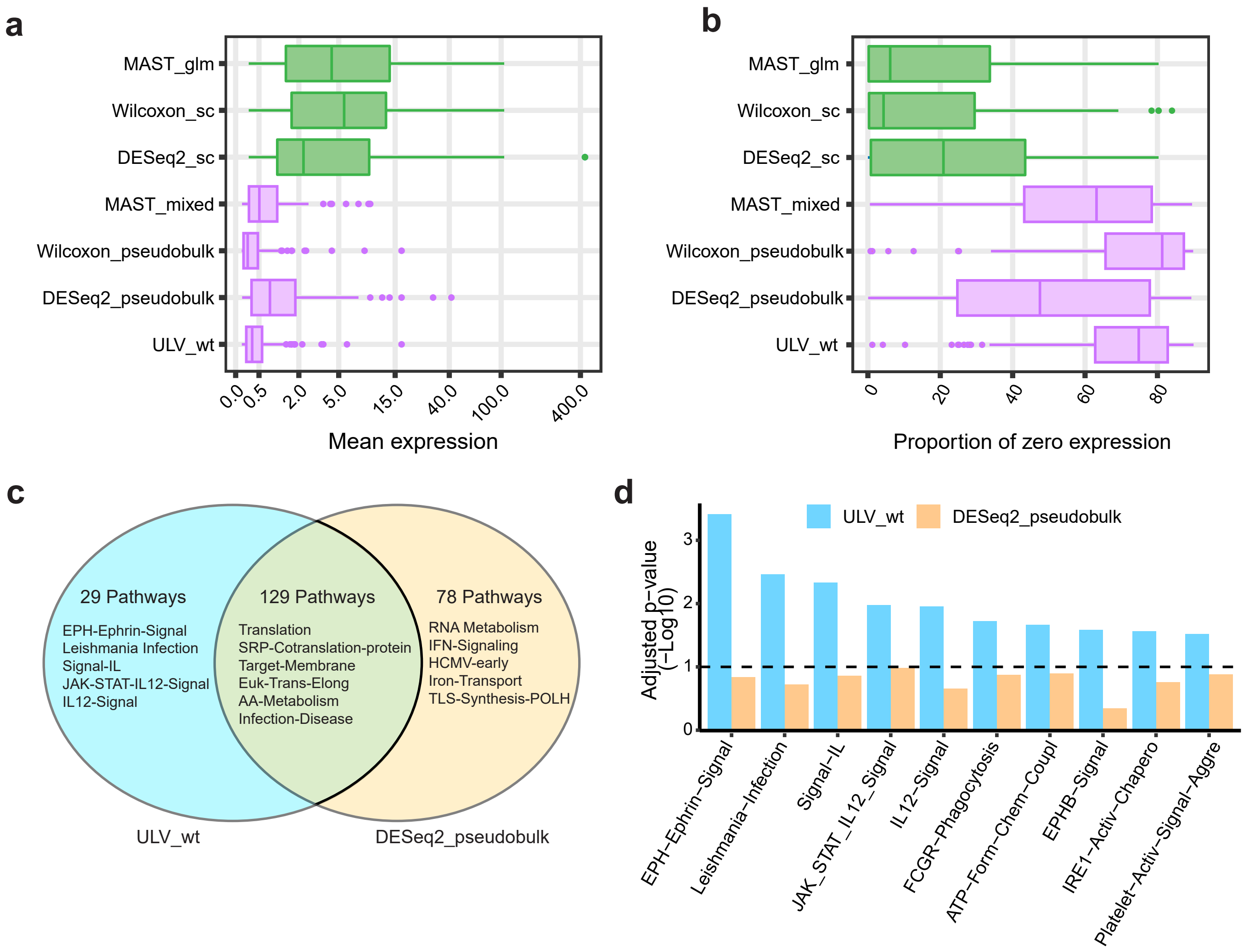

Fig 6.